\documentclass[12pt]{article}
\begin{document}
\pagestyle{plain}
\title{\bf  RADIATION OF THE GRAVITATIONAL\\[5mm]
AND ELECTROMAGNETIC BINARY \\[6mm]
PULSARS}
\author{\bf Miroslav Pardy\\[15mm]
INSTITUTE OF PLASMA PHYSICS ASCR\\
PRAGUE ASTERIX LASER SYSTEM, PALS\\
Za Slovankou 3, 182 21 Prague 8, Czech Republic\\
and\\
MASARYK UNIVERSITY\\
DEPARTMENT OF
PHYSICAL ELECTRONICS\\
Kotl\'{a}\v{r}sk\'{a} 2, 611 37
Brno, Czech Republic \\
e-mail: pamir@physics.muni.cz}
\date{\today}
\maketitle

\vspace{2cm}

\begin{abstract}
The energy-loss formula of the production of gravitons by the binary
is derived in the source theory formulation of gravity .
Then, the  quantum energy loss formula involving radiative corrections
is derived. We postulate an idea that gravitational pulsars are
present in our universe and that radiative corrections
play a role in the physics of the cosmological scale.
In the last part of the article, we consider so called electromagnetic pulsar which is
formed by two particles with the opposite electrical charges moving
in the constant magnetic field and generating the
electromagnetic pulses. We think that the cosmological analogue
is possible.
\end{abstract}
\newpage

\section{PULSARS IN GENERAL}

Pulsars are the specific cosmological objects which
radiates electromagnetic energy in the form of short pulses.
They were discovered by
Hewish et all. in 1967, published by Hewish et all. (1968) and specified
later as
neutron stars. Now, it is supposed that they are fast rotating neutron
stars with approximately Solar mass and
with strong magnetic fields ($10^9 - 10^{14}$ Gauss). The neutron
stars are formed during the evolution of stars and they
are the product of the reaction $p + e^{-} \rightarrow  n
+ \nu_{e}$, where the symbols in the last equation are as follows:
proton, electron, neutron and the electron neutrino.
The neutron stars were postulated by Landau (1932) at the
 year of the discovery of neutron in 1932. The pulsars
composed from pions, or, hyperons, or, quarks probably also exist in universe,
however it is not a measurement technique which rigorously determines
these kinds of pulsars.

Pulsars emit highly accurate periodic signals mostly in radio
waves beamed in a cone of radiation centered around their magnetic axis.
These signals define the period of rotation of the neutron
star, which radiates as a light-house once per revolution.
We know so called slow rotated pulsars, or, normal pulsars
with period P $>$ 20ms and so called millisecond pulsars
with period P $<$ 20ms.

In 1974 a pulsar in a binary system was discovered by Hulse
and Taylor and the discovery was published in 1975
(Hulse and Taylor, 1975).
The period of rotation of normal pulsars increases with time
and led to the rejection of the
suggestion that the periodic signal could be due to the orbital
period of binary stars. The orbital period of an isolated binary
system decreases as it loses energy, whereas the period of a rotating
body increases as it loses energy.

The present literature concerns only the electromagnetic pulsars.
The number of them is 1300 and they were catalogued with very precise
measurements
of their positions and rotation rates. The published pulse profiles
are so called integrated profiles obtained by adding some hundred of thousands
of individual pulses. The integration hides a large variation of size
and shape from pulse to pulse of the individual pulsar.
The radiation is emitted along the direction of the field lines,
so that the observed  duration of the integrated profile
depends on the inclination of the dipole axis to the rotation axis,
because it is supposed that the pulsar radiation is a radiation of the
dipole in the magnetic field.
It means that the radiation of these pulsars is the synchrotron
radiation of charged particles.
 We know that
the analogical ultrarelativistic charge moving in a constant magnetic
field radiates the synchrotron radiation in a very narrow
cone and such system can be considered as a
free electron laser in case that the opening angle of the cone is very
small. Of course the angle of emission of
 pulsars is not smaller than radians. In other words
the observed pulsars are not the free electron lasers. The idea that
pulsar can be a cosmical maser was also rejected.

Pulsar radio emission is highly polarized, with linear and circular
components. Individual pulses are often observed to be 100\% polarized.
The study of the polarization of pulsar is the starting point of the
determination of their real structure.

The only energy source of the pulsar is the rotational energy of the
neutron star. The rate of the dissipation of the rotational
 energy can be determined.
The moment of inertia is fairly accurately known from the
theory of the internal structure and the rotational slowdown
is very accurately measured for almost every pulsar. So most of energy
is radiated as magnetic dipole radiation at the rotation
frequency leading to a measure of the magnetic dipole and the surface
field strength.

The published articles on pulsars deal with the observation of the
pulses and  with the theoretical models. The observational results
giving an insight into the behavior of matter in the presence of
extreme gravitational and electromagnetic field are summarized for
instance by Manchester (Manchester, 1992). The emission mechanism of
photons are reviewed from a plasmatical viewpoint by Melrose (1992).
The morfology of the radio pulsars
is presented in the recent treatise by Seiradakis et. al. (2004). The
review of properties of pulsars involving the radio
propagation in the magnetosphere and of emission mechanism is
summarized  in the article by Graham-Smith (2003).
At the same time there are, to our knowledge,
no information on so called gravitational
pulsars, or, on the models where the pulses are
produced by the retrograde motion of the bodies moving around the
central body.

So,  the question arises, if it is possible to define gravitational
binary pulsar, where the gravitational
energy is generated by the binary
system, or, by the system where two components are in the retrograde motion.
 We suppose  that in case of the massive
 binary system the energy is generated in a cone starting from the
component of a binary and it can be seen
only if the observer is present  in the axis this cone.
Then, the observer detects gravitational pulses when  the
detector is sufficiently sensitive. There are many methods for the
detection of the gravitational waves. One method, based on the quantum
states of the superfluid ring, was suggested  by author (Pardy, 1989).

We know that gravitational waves was indirectly confirmed by the
observation of the period of the pulsar PSR 1913 + 16.
The energy loss of this pulsar was calculated in the
framework of the classical theory of gravitation. The quantum energy
loss  was given for instance
by Manoukian (1990). His calculation  was based
on the so called Schwinger source theory where gravity is
considered as a field theory of gravitons
 where graviton is a boson with spin 2, helicity $\pm$2  and zero mass.
It is an analogue of photon in the electromagnetic theory.

In the following text we start with the source derivation of the
power spectral formula of the
gravitational radiation of a binary.
Then, we calculate the quantum energy loss of a binary and
the gravitational power spectrum involving
radiative corrections.
In the last part of an article,
we consider so called electromagnetic pulsar which is
formed by two particles
with the opposite electrical charges  which
move in the constant magnetic field and generate the
electromagnetic pulses.

\section{THE QUANTUM GRAVITY ENERGY LOOS OF A BINARY SYSTEM}

\subsection{Introduction}

At the present time, the existence of gravitational waves is
confirmed, thanks to the experimental proof of Taylor
and Hulse who performed the systematic measurement of the motion of the binary
with the pulsar PSR 1913+16. They found that the generalized
energy-loss formula, which follows
from the Einstein general theory of relativity, is in accordance with
their measurement.

This success was conditioned by the fact that the binary with the
pulsar PSR 1913+16 as a gigantic system of two neutron stars,
emits sufficient gravitational radiation to influence the orbital
motion of the binary at the observable scale.

Taylor and Hulse, working at the Arecibo radiotelescope, discovered
the radiopulsar PSR 1913+16 in a binary, in 1974, and
this is now considered as the best general relativistic laboratory (Taylor,
1993).

Pulsar PSR 1913+16 is the massive body of the binary system where
each of the rotating pairs is 1.4 times the mass of the Sun. These
neutron stars rotate around each other
with a period 7.8 hours, in an orbit not much larger
than the Sun's diameter. Every 59 ms,
the pulsar emits a short signal that is so clear that the arrival
time of a 5-min string of a set of such signals can be resolved
within 15 $\mu$s.

A pulsar model based on strongly magnetized, rapidly spinning neutron
stars was soon established as consistent with most of the known
facts (Huguenin et al, 1968); its electrodynamical properties
were studied theoretically
(Gold, 1968) and shown to be plausibly capable of generating
broadband radio noise
detectable over interstellar distances. The binary pulsar PSR 1913+16
is now recognized as the harbinger of a new class of unusually
short-period pulsars, with numerous important applications.

Because the velocities and
gravitational energies in a high-mass binary pulsar system can be
significantly relativistic, strong-field and radiative effects
come into play. The binary pulsar PSR 1913+16 provides significant tests of
gravitation beyond the weak-field, slow-motion limit (Goldreichet al., 1969;
Damour et al., 1992).

The goal of this section  is not to repeat the derivation of the Einstein
quadrupole formula, because this has been performed many
times in general relativity
and also in the Schwinger source theory in the weak-field limit
(Manoukian, 1990).

We show that just in the framework of the source theory it is
easy to determine the quantum energy-loss formula
of the binary system. The energy-loss formula can be generalized
in such a way it involves also the radiative corrections.

Since the measurement of the motion of the binaries goes on,
we hope that future experiments will verify the quantum version
of the energy-loss formula, involving also the radiative corrections.

\subsection{The source theory  formulation  of the problem}

We show how the total quantum loss of energy caused
by the production of gravitons, emitted
by the binary system of two point masses moving around
each other under their gravitational interaction, can be  calculated
in the framework of the source theory of gravity.

Source theory (Schwinger, 1970, 1973, 1976) was initially constructed
to describe the particle physics situations occurring in
high-energy physics experiments. However, it was found that the
original formulation simplifies the calculations in the
electrodynamics and gravity, where the interactions are mediated by
photon and graviton respectively. The source theory of
gravity forms the analogue of quantum electrodynamics because, while in
QED the interaction is mediated by the photon, the gravitational interaction is
mediated by the graviton (Schwinger, 1976).

The source theory of gravity invented by Schwinger is linear theory.
So, the question arises
 if it is in the coincidence with the Einstein gravity equations
which are  substantially nonlinear.
 The answer is affirmative, because the coincidence is
only with the linear approximation of the Einstein theory.
The experimental results of the Schwinger theory are also
in harmony with experiment.
The quadrupole formula of Einstein also follows from the
Schwinger version.

The unification of gravity and electromagnetism is possible only in
the Schwinger source theory.
It is possible when, and only when it is  possible the
unification of forces. And this is performed in the Schwinger
source theory of all interactions where force is of the Yukawa form.
The problem of unification is not new.
We know from the history of physics that the Ptolemy system
could not be unified with the Galileo-Newton system
because in the Ptolemy system it is not defined the force which is the
fundamental quantity in the GN system and the primary cause of all
phenomena in this system.

Einstein gravity uses the Riemann space-time where the gravity force
has  not the Yukawa dynamical form.
The curvature of space-time is defined as the origin of all
phenomena. The gravity force in the Einstein theory is in the antagonistic
contradistinction with the Yukawa force in the quantum field theory
and therefore it seems that
QFT, QED, QCD and EL.-WEAK theory cannot be unified with the Einstein gravity.

Manoukian (1990) derivation of
the Einstein quadrupole formula in the framework of the Schwinger
source theory is possible because of the coincidence of the source theory
with the linear limit of the Einstein theory.

Our approach is different from Manoukian method because we derive the
power spectral spectral formula $ P(\omega)$
of emitted gravitons with frequency $\omega$  and then using
relation $(- dE/dt) = \int d\omega P(\omega)$ we determine  the energy
loss $E$.
In case of the radiative correction, we derive only the
power spectral formula in the general form.

The mathematical structure of $P(\omega )$ follows directly
from the action $W$ and while in the case of the gravitational radiation,
the formula is composed from the tensor of  energy-momentum,
then, in the case of the electromagnetic radiation  formula, it  involves
charged vector currents.

The basic formula in the source
theory is the vacuum-to-vacuum amplitude (Schwinger et al., 1976):

$$
\langle 0_{+}|0_{-}\rangle = e^{\frac {i}{\hbar}\*W(S)},\eqno(1)$$
where the minus and plus tags on the vacuum symbol are causal
labels, referring to any time before and after region of space-time,
where sources are manipulated. The exponential form is introduced
with regard to the existence of the physically independent
experimental arrangements, which has the simple consequence that the
associated probability amplitudes multiply and the corresponding
$W$ expressions add (Schwinger et al., 1976; Dittrich, 1978).

In the flat space-time, the field of gravitons is described by
the amplitude (1) with the action (Schwinger, 1970)
($c = 1$ in the following text)

$$W(T) = 4\pi\*G\*\int (dx)(dx')$$

$$\times \quad \left[T^{\mu\nu}(x)
\*D_{+}(x-x')T_{\mu\nu}(x')
 -  \frac{1}{2}\*T(x)D_{+}(x-x')T(x')\right],\eqno(2)$$
where the dimensionality of $W(T)$ is the same as the dimensionality
of the Planck constant $\hbar$; $T_{\mu\nu}$ is the tensor of momentum and
energy. For a particle moving along the trajectory
${\mbox {\bf x}} = {\mbox {\bf x}}(t)$, it is defined by the equation
(Weinberg, 1972):

$$T^{\mu\nu}(x) = \frac{p^{\mu}p^{\nu}}{E}\*\delta({\mbox {\bf x}} -
{\mbox {\bf x}}(t)), \eqno(3)$$

\noindent
where $p^{\mu}$ is the relativistic four-momentum of a particle
with a rest mass $m$  and

$$p^{\mu} = (E,{\mbox {\bf p}}) \eqno(4)$$

$$p^{\mu}\*p_{\mu} = - m^2,\eqno(5)$$

\noindent
and the relativistic energy is defined by the known relation

$$ E = \frac {m}{\sqrt{1 - {\mbox {\bf v}}^{2}}},\eqno(6)$$

\noindent
where {\mbox {\bf v}} is the three-velocity of the moving particle.

Symbol $T(x)$ in formula (2) is defined as
$T = g_{\mu\nu}T^{\mu\nu}$, and $D_{+}(x-x')$ is the graviton
propagator whose explicit form will be determined later.

The action $W$ is not arbitrary because it must involve
the attractive  force between the gravity masses
while in case of the electromagnetic situation the action must
involve the repulsive force between
charges of the same sign. It is very surprising that such form of
Lagrangians follows  from the
quantum definition of the vacuum to vacuum amplitude. It was shown
by Schwinger that Einstein
gravity also follows from the source theory, however the method of
derivation is not the integral part of the source
theory because the source theory is linear and it is not clear how
to establish the equivalence
between linear and nonlinear theory. String theory tries to solve
the problem of the unification of all forces,
however, this theory is, at the present time,  not predictable and works
with so called extra-dimensions which was not observed.
It is not clear from the viewpoint of physics, what the dimension is.
It seems that many
problems can be solved in the framework of the source theory.

\subsection{The power spectral formula in general}

It may be easy to show that the probability of the persistence
of vacuum is given by the following formula (Schwinger et al., 1976):

$$|\langle 0_{+}|0_{-}\rangle|^2 =
\exp\left\{-\frac {2}{\hbar}{\mbox {\rm Im}}\* W\right\} \,
\stackrel{d}{=}\, \exp\left\{-\int\,dtd\omega
\frac {1}{\hbar\omega}P(\omega,t)\right\},\eqno(7)$$

\noindent
where the so-called power spectral function $P(\omega,t)$ has been introduced
(Schwinger et al., 1976). In order to extract this
spectral function from Im $W$, it
is necessary to know the explicit form of the graviton propagator
$D_{+}(x-x')$. The physical content of this propagator is
analogous to the content of the photon propagator. It involves
the gravitons property
of spreading with velocity $c$. It means that
its explicit form is just the same as that of the photon propagator.
With regard to the source theory
(Schwinger et al., 1976) the $x$-representation
of $D_{+}(x)$ in eq. (2) is as follows:

$$D_{+}(x-x') = \int \frac {(dk)}{(2\pi)^4}\*e^{ik(x-x')}\*D(k),\eqno(8)$$

\noindent
where

$$ D(k) = \frac {1}{|{\mbox {\bf k}}^2| - (k^0)^2 - i\epsilon},\eqno(9)$$

\noindent
which gives
$$ D_{+}(x-x') =\frac {i}{4\pi^2}\*\int_{0}^{\infty}d\omega
\frac {\sin\omega|{\mbox {\bf x}}-{\mbox {\bf x}}'|}
{|{\mbox {\bf x}} - {\mbox {\bf x}}'|}\* e^{-i\omega|t-t'|}. \eqno(10)$$

Now, using  formulas (2), (7) and (10), we get the power spectral formula
in the following form:

$$P(\omega,t) = 4\pi\*G\*\omega\int\, (d{\mbox {\bf x}})
(d{\mbox {\bf x}}')dt'\*
\frac {\sin\omega|{\mbox {\bf x}}-{\mbox {\bf x}}'|}
{|{\mbox {\bf x}}-{\mbox {\bf x}}'|}\*\cos\omega(t- t') $$

$$\times \quad
\left[T^{\mu\nu}({\mbox {\bf x}},t)T_{\mu\nu}
({\mbox {\bf x}}',t') - \frac{1}{2}g_{\mu\nu}T^{\mu\nu}({\mbox {\bf x}},t)
g_{\alpha\beta}T^{\alpha\beta}({\mbox {\bf x}}',t')\right].
\eqno(11)$$

\subsection{The power spectral formula for the binary system}

In the case of the binary system with masses $m_{1}$ and $m_{2}$, we
suppose that they move in a uniform circular motion around their center
of gravity in the $xy$ plane, with corresponding kinematical coordinates:

$${\mbox {\bf x}}_{1}(t) =
r_{1}({\mbox {\bf i}}\cos(\omega_{0}t) + {\mbox {\bf
    j}}\sin(\omega_{0}t))
\eqno(12)$$

$${\mbox {\bf x}}_{2}(t) =  r_{2}({\mbox {\bf i}}\cos(\omega_{0}t + \pi) +
{\mbox {\bf j}}\sin(\omega_{0}t + \pi))\eqno(13)$$

\noindent
with
$${\mbox {\bf v}}_{i}(t) = d{\mbox {\bf x}}_{i}/dt, \hspace{5mm} \omega_{0} =
v_{i}/r_{i}, \hspace{5mm} v_{i} = |{\mbox {\bf v}}_{i}|
\quad (i = 1,\, 2). \eqno(14)$$

For the tensor of energy and momentum of the binary we have:

$$T^{\mu\nu}(x) =
\frac{p_{1}^{\mu}p_{1}^{\nu}}{E_{1}}\*\delta({\mbox {\bf x}} -
{\mbox {\bf x}}_{1}(t)) +
\frac{p_{2}^{\mu}p_{2}^{\nu}}{E_{2}}\*\delta({\mbox {\bf x}} -
{\mbox {\bf x}}_{2}(t)),
\eqno(15)$$
where we have omitted the tensor $t^G_{\mu\nu}$, which is
associated with the massless,
gravitational field distributed all over space and proportional to
the gravitational constant $G$ (Cho et al., 1976):

After insertion of eq. (15) into eq. (11), we get:

$$P_{total}(\omega,t) = P_{1}(\omega,t) +P_{12}(\omega,t) + P_{2}(\omega,t),
\eqno(16)$$
where ($t' - t = \tau$):

$$P_{1}(\omega,t) = \frac {G\omega}{r_{1}\pi}\* \int_{-\infty}^{\infty}\,
d\tau\*\frac {\sin[2\omega\*r_{1}\*\sin(\omega_{0}\tau/2)]}
{\sin(\omega_{0}\tau/2)}\*\cos\omega\tau$$

$$\times\quad \left(E_{1}^2(\omega_{0}^2\*r_{1}^2\*\cos\omega_{0}\tau - 1)^2 -
\frac {m_{1}^4}{2E_{1}^2}\right),\eqno(17)$$

$$P_{2}(\omega,t) = \frac {G\omega}{r_{2}\pi}\* \int_{-\infty}^{\infty}\,
d\tau\*\frac {\sin[2\omega\*r_{2}\*\sin(\omega_{0}\tau/2)]}
{\sin(\omega_{0}\tau/2)}\*\cos\omega\tau$$

$$\times\quad \left(E_{2}^2(\omega_{0}^2\*r_{2}^2\*\cos\omega_{0}\tau - 1)^2 -
\frac {m_{2}^4}{2E_{2}^2}\right),\eqno(18)$$

$$P_{12}(\omega,t) = \frac {4G\omega}{\pi}\* \int_{-\infty}^{\infty}\,
d\tau\*\frac {\sin\omega\*[r_{1}^2 + r_{2}^2
+ 2r_{1}\*r_{2}\cos(\omega_{0}\tau)]^{1/2}}
{[r_{1}^2 + r_{2}^2 +  2r_{1}\* r_{2}\* \cos(\omega_{0}\tau)]^{1/2}}
\*\cos\omega\tau$$

$$\times\quad \left(E_{1}\*E_{2}(\omega_{0}^2\*r_{1}\*r_{2}
\*\cos\omega_{0}\tau + 1)^2 -
\frac {m_{1}^2\*m_{2}^2}{2E_{1}\*E_{2}}\right) .\eqno(19)$$

\subsection{The quantum energy loss of the binary}

Using the following relations

$$\omega_{0}\tau = \varphi + 2\pi\*l,  \hspace{7mm} \varphi\in(-\pi,\pi),
\quad l = 0,\, \pm1,\, \pm2,\, ...\eqno(20)$$

$$\sum_{l=-\infty}^{l=\infty}\; \cos2\pi\*l \frac {\omega}{\omega_{0}} =
\sum_{l=-\infty}^\infty \; \omega_{0}\delta(\omega - \omega_{0}l), \eqno(21)$$
we get for $P_{i}(\omega,t)$, with $\omega$ being restricted to positive:

$$P_{i}(\omega,t) = \sum_{l=1}^{\infty}\;\delta(\omega-\omega_{0}\*l)\*
P_{il}(\omega,t).\eqno(22)$$

Using the definition of the Bessel function $J_{2l}(z)$

$$J_{2l}(z) = \frac {1}{2\pi}\int_{-\pi}^{\pi}\,d\varphi\*
\cos \left(z\sin\frac{\varphi}{2}\right)\*\cos{l\varphi}, \eqno(23)$$

\noindent
from which the derivatives and their integrals follow, we get for
$P_{1l}$ and $P_{2l}$ the following formulas:

$$P_{il} = \frac {2G\omega}{r_{i}}\* \Bigl((E_{i}^2(1 - v_{i}^2)^{2} -
\frac {m_{i}^4}{2E_{i}^2}\Bigr)\*\int_{0}^{2v_{i}\*l}\,dx\,J_{2l}(x)
$$

$$ + \quad 4E_{i}^2(1 - v_{i}^2)\*v_{i}^2\*J'_{2l}(2v_{i}l) +
4E_{i}^2v_{i}^4\*J'''_{2l}(2v_{i}l)\Bigr),\quad i = 1,\, 2. \eqno(24)$$

Using $r_{2} = r_{1} + \epsilon $,
where $\epsilon$ is supposed to be small in comparison with radii
$r_{1}$ and $r_{2}$, we obtain

$$[r_{1}^2 + r_{2}^2 + 2r_{1}\*r_{2}\cos\varphi]^{1/2} \approx
2a\cos\left(\frac {\varphi}{2}\right), \eqno(25)$$
with

$$a = r_{1}\left(1 + \frac {\epsilon}{2r_{1}}\right). \eqno(26)$$

So, instead of eq. (19) we get:

$$P_{12}(\omega,t) = \frac {2G\omega}{a\pi}\* \int_{-\infty}^{\infty}\,
d\tau\*\frac {\sin[2\omega\*a\*\cos(\omega_{0}\tau/2)]}
{\cos(\omega_{0}\tau)/2]}
\*\cos\omega\tau$$

$$\times\quad
\left(E_{1}\*E_{2}(\omega_{0}^2\*r_{1}\*r_{2}\*\cos\omega_{0}\tau + 1)^2 -
\frac {m_{1}^2\*m_{2}^2}{2E_{1}\*E_{2}}\right).\eqno(27)$$

Now, we can approach the evaluation of the energy-loss formula
for the binary from the power spectral formulas (24) and (27).
The energy loss is defined by the relation

$$-\frac{dU}{dt} = \int\,P(\omega)d\omega = $$
$$\int\,d\omega\*\sum_{i,l}\delta(\omega - \omega_{0}l)P_{il}
+ \int\, P_{12}(\omega)d\omega =
-\frac {d}{dt}(U_{1} + U_{2} + U_{12}). \eqno(28)$$
Or,

$$-\frac {d}{dt}U_{i} =
\int\,d\omega\*\sum_{l}\delta(\omega - \omega_{0}l)P_{il},\quad
-\frac {d}{dt}U_{12} =
\int\,d\omega\*\sum_{i,l}\delta(\omega - \omega_{0}l)P_{12l}.\eqno(29)
$$

From Sokolov and Ternov (1983) we learn  Kapteyn's formulas:

$$\sum_{l=1}^{\infty} 2l\*J'_{2l}(2lv) = \frac {v}{(1 - v^2)^2},
 \eqno(30)$$
and

$$\sum_{l=1}^{\infty} l\*\int_{0}^{2lv}\*J_{2l}(x)dx =
\frac {v^3}{3(1-v^2)^3}. \eqno(31)$$

The formula $\sum_{l=1}^{\infty} l\*J'''_{2l}(2lv) = 0$ can be obtained
from formula

$$\sum_{l=1}^{\infty} \frac {1}{l}\*J'_{2l}(2lv) = \frac{1}{2}v^{2}
 \eqno(32)$$
by its differentiation with the respect to $v$ (Schott, 1912).

Then,  after application of eqs. (30), (31) and (32) to eqs.
(24) and (28), we get:

$$-\frac {dU_{i}}{dt} =
\frac {2G\omega_{0}}{r_{i}}\*
\left[\left(E_{i}^{2}( v_{i}^2 -1)^{2} -
 \frac{m_{i}^{4}}{2E_{i}^{2}}\right)\frac{v_{i}^{3}}{3(1 - v_{i}^{2})^{3}} -
2E_{i}^{2}v_{i}^{3} + 4E_{i}^{2}v_{i}^{4}\right]. \eqno(33)$$

Instead of using Kapteyn's formulas for the interference term,
we will perform a direct evaluation
of the energy loss of the interference term by the $\omega$-integration
in  (27). So, after some elementary modification in the $\omega$-integral,
we get:

$$- \frac {dU_{12}}{dt} =  \int_{0}^{\infty}\,P(\omega)d\omega = $$

$$A\*\int_{-\infty}^{\infty}d\tau\*\int_{-\infty}^{\infty}\,d\omega\*\omega
\*e^{-i\omega\tau}\*\sin[2\omega\*a\*\cos \omega_{0}\tau]\*
\left[\frac {B(C\cos\omega_{0}\tau + 1)^2 - D}{\cos(\omega_{0}\tau/2)}
\right],
\eqno(34)$$
with

$$A = \frac {G}{a\pi},\quad B = E_{1}E_{2},\quad C = v_{1}v_{2},
\quad D = \frac {m_{1}^2\*m_{2}^2}{2E_{1}E_{2}}.\eqno(35)$$

Using the definition of the $\delta$-function and its derivative,
we have, instead of eq. (34), with $v = a\omega_{0}$:

$$- \frac {dU_{12}}{dt} =
A\*\omega_{0}\pi\*\int_{-\infty}^{\infty}\,dx\,
\*\frac {[B(C\*\cos x + 1)^{2} - D]}{\cos(x/2)}\quad  \times $$

$$\left[\delta'(x - 2v\cos(x/2)) - \delta'(x + 2v\cos(x/2))\right].
\eqno(36)$$

Putting

$$x - 2v\cos(x/2) = t, \eqno(37)$$
in the first $\delta'$-term and

$$y + 2v\cos(y/2) = t, \eqno(38)$$
in the second  $\delta'$-term, we get eq. (36) in the following form:

$$-\frac{dU_{12}}{dt} = 2A\omega_{0}v\pi \*
\int_{-\infty}^{\infty}\,dt\delta'(t) \times$$

$$\left\{\frac {[B(C\cos x + 1)^2 - D]}{(x-t)(1+v\sin(x/2))} -
\frac {[B(C\cos y + 1)^2 - D]}{(y+t)(1-v\sin(y/2))}.
\right\} \eqno(39)$$

Using the known relation for a $\delta$-function:

$$\int\,dtf(t)\delta'(t) = - f'(0), \eqno(40)$$
we get the energy loss formula for the synergic term in the form:

$$-\frac{dU_{12}}{dt} = -2A\omega_{0}v\pi \* \left.
\frac{d}{dt}\left\{\frac {[B(C\cos x + 1)^2 - D]}{(x-t)(1+v\sin(x/2))} -
\frac {[B(C\cos y + 1)^2 - D]}{(y-t)(1-v\sin(y/2))}
\right\}\right |_{t = 0}, \eqno(41)$$
and we recommend the final calculation of the last formula to
the mathematical students.

Let us remark finally that the formulas derived
for the energy loss of the binary
(33) and (41) describes only the binary system and therefore
their sum  has not the form of the Einstein quadrupole formula.
The sum forms the total produced gravitational energy,
and involves not only the radiation of the individual bodies of the
binary, but also the interference term.
The problem of the coincidence
with the Einstein quadrupole formula is open.

\section{THE POWER SPECTAL FORMULA IVOLVING RADIATIVE CORRECTIONS}

\subsection{Introduction}

We here calculate the total quantum loss of energy caused by production of gravitons emitted
by the binary system in the framework of the source
theory of gravity for the situation with the gravitational
propagator involving radiative corrections.

We know from  QED that photon can exist in the virtual state as
the two body system in the form of the
electron positron pair. It means that the photon propagator involves
the additional process:

$$\gamma \rightarrow e^{+} + e^{-} \rightarrow \gamma .\eqno(42)$$

In case of the graviton radiation, the situation is analogical with
 the situation in QED. Instead of eq. (42) we write

$$g \rightarrow 2e^{+} + 2e^{-} \rightarrow g ,\eqno(43)$$
where $g$ is graviton  and number 2 is there in order to conserve spin also during the virtual process.

Equation (43) can be of course expressed in more detail:

$$g \rightarrow \gamma + \gamma \rightarrow (e^{+} + e^{-}) + (e^{+} + e^{-})
 \rightarrow \gamma  + \gamma  \rightarrow  g.   \eqno(44)$$

We will show that in the framework of the source theory it is
easy to determine the quantum energy loss formula
of the binary system both in case  with the graviton
propagator with radiative corrections.

We will investigate how the spectrum of the  gravitational
radiation is
modified if we involve radiation corrections corresponding to the
virtual pair production and annihilation in the graviton
propagator. Our calculation is an analogue of the
photon propagator with radiative corrections for production of photons
by the \v{C}erenkov mechanism (Pardy, 1994c,d).

Because the measurement of motion of the binaries goes on,
we hope that the future experiments will verify the quantum version
of the energy loss formula following from the source theory
and that sooner or later the confirmation of this formula will be established.

\subsection{The binary power spectrum with radiative corrections}

According to source theory (Schwinger, 1973;
Dittrich, 1978; Pardy, 1994c,d), the photon propagator in the Minkowski space-time
with radiative correction is in the momentum representation of the form:

$$\tilde{D}(k) = D(k) + \delta D(k), \eqno(45)$$
or,

$$\tilde{D}(k) = \frac {1}{|{\mbox {\bf k}}|^2-(k^0)^2-i\epsilon} $$

$$+ \quad \int_{4m^2}^\infty dM^2 \frac {a(M^2)}
{|{\mbox {\bf k}}|^2-(k^0)^2+\frac {M^2\*c^2}{\hbar^{2}}-i\epsilon}, \eqno(46)$$
where $m$ is mass of electron and
the last term in equation (44) is derived on the virtual
photon condition

$$|{\mbox {\bf k}}|^2 - (k^0)^2 = - \frac {M^2\*c^2}{\hbar^{2}}.\eqno(47)$$

The weight function $a(M^2)$ has been derived in the following form
(Schwinger, 1973; Dittrich, 1976):

$$a(M^2) = \frac {\alpha}{3\pi} \frac {1}{M^2} \left(1+\frac {2m^2}{M^2}\right)
\left(1 - \frac {4m^2}{M^2}\right)^{1/2}.  \eqno(48)$$

We suppose that the graviton propagator with the radiative corrections
forms the analogue of the photon propagator.

Now, with regard to the definition of the Fourier transform
$$
D_{+}(x-x') = \int \frac {(dk)}{(2\pi)^4}\* e^{ik(x-x')}\*D(k),\eqno(49)
$$
we get for $\delta\*D_{+}$ the following relation ($c = \hbar = 1$):

$$\delta\*D_{+}(x-x') = \frac {i}{4\pi^2}
\*\int_{4m^2}^{\infty}\,dM^2\*a(M^2)$$

$$\times\; \int\,d\omega\,\frac {\sin\left\{[\omega^{2}-
M^2]^{1/2}\*
|{\mbox {\bf x}}-{\mbox {\bf x}'}|\right\}}{|{\mbox {\bf x}}-
{\mbox {\bf x}}'|}\*e^{-i\omega\*|t-t'|}.\eqno(50)$$

The function (50) differs from the gravitational function
"$D_{+}$" in (9) especially by the factor

$$\left(\omega^2 - M^2 \right)^{1/2}\eqno(51)$$

\noindent
in the function '$\sin$'  and by the additional mass-integral which
involves the radiative corrections to the original power spectrum formula.

In order to determine the additional spectral function of produced gravitons,
corresponding to the radiative corrections, we insert
$D_{+}(x-x') + \delta D_{+}(x-x')$ into eq. (2), and using eq. (11) we
obtain (factor 2 from the two photons is involved):

$$\delta\*P(\omega,t) =
\frac {4\*G\omega}{\pi}\int\, (d{\mbox {\bf x}})(d{\mbox {\bf x}}')dt'\*
\int_{4m^2}^{\infty}\,dM^2\*a(M^2)$$

$$\times\quad \frac {\sin\left\{[\omega^2 -
M^2]^{1/2}|{\mbox {\bf x}}-{\mbox {\bf x}}'|\right\}}
{|{\mbox {\bf x}}-{\mbox {\bf x}}'|}\*\cos\omega(t- t') $$

$$\times \quad
\Bigl [T^{\mu\nu}({\mbox {\bf x}},t)g_{\mu\alpha}g_{\nu\beta}T^{\alpha\beta}
({\mbox {\bf x}}',t') - \frac{1}{2}g_{\mu\nu}T^{\mu\nu}({\mbox {\bf x}},t)
g_{\alpha\beta}T^{\alpha\beta}({\mbox {\bf x}}',t')\Bigr].\eqno(52)$$

Then using eqs. (16), (17), (18) and (19), we get

$$\delta P_{total}(\omega,t) =\delta  P_{1}(\omega,t) + \delta  P_{2}(\omega,t) + \delta  P_{12}(\omega,t),
\eqno(53)$$
where ($t' - t = \tau$):

$$\delta  P_{1}(\omega,t) =
\frac {2G\omega}{r_{1}\pi}\* \int_{-\infty}^{\infty}\,
d\tau\*\int_{4m^2}^{\infty}\,dM^2\*a(M^2)
\frac {\sin\{2\left(\omega^2 - M^2 \right)^{1/2}\*r_{1}\*\sin(\omega_{0}\tau/2)\}}
{\sin(\omega_{0}\tau/2)}\*\cos\omega\tau $$

$$\times \quad
\left(E_{1}^2(\omega_{0}^2\*r_{1}^2\*\cos\omega_{0}\tau - 1)^2 -
\frac {m_{1}^4}{2E_{1}^2}\right),\eqno(54)$$

$$\delta  P_{2}(\omega,t) = \frac {2G\omega}{r_{2}\pi}\*
\int_{-\infty}^{\infty}\,
d\tau\*\int_{4m^2}^{\infty}\,dM^2\*a(M^2)
\frac {\sin\{2\left(\omega^2 - M^2
  \right)^{1/2}\*r_{2}\*\sin(\omega_{0}
\tau/2)\}}
{\sin(\omega_{0}\tau/2)}\*\cos\omega\tau $$

$$\times \quad
\left(E_{2}^2(\omega_{0}^2\*r_{2}^2\*\cos\omega_{0}\tau - 1)^2 -
\frac {m_{2}^4}{2E_{2}^2}\right),\eqno(55)$$

$$\delta  P_{12}(\omega,t) = \frac {8G\omega}{\pi}\* \int_{-\infty}^{\infty}\,
d\tau\*\int_{4m^2}^{\infty}\,dM^2\*a(M^2)$$

$$\frac {\sin\{\left(\omega^2 - M^2 \right)^{1/2}\*[r_{1}^2 + r_{2}^2
+ 2r_{1}\*r_{2}\cos(\omega_{0}\tau)]^{1/2}\}}
{[r_{1}^2 + r_{2}^2 +  2r_{1}\* r_{2}\* \cos(\omega_{0}\tau)]^{1/2}}
\*\cos\omega\tau $$

$$\times \quad
\left(E_{1}\*E_{2}(\omega_{0}^2\*r_{1}\*r_{2}\*\cos\omega_{0}\tau + 1)^2 -
\frac {m_{1}^2\*m_{2}^2}{2E_{1}\*E_{2}}\right).\eqno(56)$$

The explicit determination of the power spectrum is the problem which
was solved by author
in 1994. The solution was performed only approximately. Here also can
be expected only the approximative solution.
From this solution can be then derived the energy loss as in the
previous article.

Let us show the possible way of the determination of the spectral
formula. If we introduce the new variable $s$ by the relation

$$\omega^{2} - M^{2} = s^{2};  \quad  -dM^{2} = 2sds \eqno(57)$$
then, instead of eqs. (54), (55) and (56) we have

$$\delta  P_{1}(\omega,t) = \frac {2G\omega}{r_{1}\pi}
\* \int_{-\infty}^{\infty}\,
d\tau\*\int_{s_{1}}^{s_{2}}\,(2sds)\*a(\omega^{2}- s^{2})
\frac {\sin\{2 s\*r_{1}\*\sin(\omega_{0}\tau/2)\}}
{\sin(\omega_{0}\tau/2)}\*\cos\omega\tau $$

$$ \times \quad
\left(E_{1}^2(\omega_{0}^2\*r_{1}^2\*\cos\omega_{0}\tau - 1)^2 -
\frac {m_{1}^4}{2E_{1}^2}\right),\eqno(58)$$

$$\delta  P_{2}(\omega,t) = \frac {2G\omega}{r_{2}\pi}
\* \int_{-\infty}^{\infty}\,
d\tau\*\int_{s_{1}}^{s_{2}}\,(2sds)\*a(\omega^2 -s^{2})
\frac {\sin \{2s\*r_{2}\*\sin(\omega_{0}\tau/2)\}}
{\sin(\omega_{0}\tau/2)}\*\cos\omega\tau $$

$$\times \quad
\left(E_{2}^2(\omega_{0}^2\*r_{2}^2\*\cos\omega_{0}\tau - 1)^2 -
\frac {m_{2}^4}{2E_{2}^2}\right),\eqno(59)$$

$$\delta  P_{12}(\omega,t) = \frac {8G\omega}{\pi}\* \int_{-\infty}^{\infty}\,
d\tau\*\int_{s_{1}}^{s_{2}}\,(2sds)\*a(\omega^2 - s^{2})
\frac {\sin\{2 s \*[r_{1}^2 + r_{2}^2
+ 2r_{1}\*r_{2}\cos(\omega_{0}\tau)]^{1/2}\}}
{[r_{1}^2 + r_{2}^2 +  2r_{1}\* r_{2}\* \cos(\omega_{0}\tau)]^{1/2}}
 $$

$$\times \quad\*\cos\omega\tau
\left(E_{1}\*E_{2}(\omega_{0}^2\*r_{1}\*r_{2}\*\cos\omega_{0}\tau + 1)^2 -
\frac {m_{1}^2\*m_{2}^2}{2E_{1}\*E_{2}}\right),\eqno(60)$$
where

$$s_{1} = \omega^{2}  - 4m^{2}, \quad  s_{2} = \infty .\eqno(61)$$

It seems that the rigorous procedure is the $\tau$-integration as
the first step and then $s$-integration as the  second step
(Pardy, 1994c,d). While, in case of the linear motion the mathematical
operations are easy (Pardy, 1994c), in case of the circular
motion there are some difficulties (Pardy, 1994d). The final form of
eq. (60) is recommended for the mathematical experts.

The energy loss is given as follows:

$$-\frac{dU_{i}}{dt} = \int_{0}^{\infty}d\omega P_{i}(\omega, t);\quad
-\frac{dU_{12}}{dt} = \int_{0}^{\infty}d\omega P_{12}(\omega,t).
\eqno(62) $$

Now, let us go to
the discussion on the electromagnetic system of the two opposite
charges moving in the constant magnetic field and producing the pulse
synchrotron radiation.

\section{ELECTROMAGNETIC PULSAR}

\subsection{Introduction}

Here, the power spectrum formula of the synchrotron
radiation generated by the
electron and positron moving at the opposite angular
velocities in homogeneous magnetic field
is derived in the Schwinger version of quantum field theory.

It is surprising that the spectrum
depends periodically on radiation frequency $\omega$ and time which means
that the system composed from electron, positron and magnetic
field forms the pulsar.

We will show that the large hadron collider (LHC) which is at
present time under construction in CERN  can be considered in near future
also as the largest electromagnetic terrestrial pulsar.
We know that while the Fermilab's Tevatron handles counter-rotating
protons and antiprotons in a single beam channel, LHC  will operate
with proton and proton  beams in such a way that the
collision center of mass energy will be 14 TeV and luminosity $10^{34} {\rm
cm}^{-1} {\rm s}^{-2}$. To achieve such large luminosity it must operate
with more than 2800 bunches per beam and a very high density of particles in
bunches. The LHC will also operate for heavy Pb ion physics at a luminosity
of $10^{27} {\rm cm}^{-1} {\rm s}^{-2}$ (Evans, 1999).

The collision of particles is caused by the opposite directional
motion of bunches.
Or, if one bunch has the angular velocity $\omega$, then the bunch with
antiparticles has angular velocity $-\omega$. Here we will determine
the spectral density of emitted photons in the simplified case
where one electron and one positron move in the opposite direction on a circle.
We will show that the synergic
spectrum depends  periodically on time. This means that the
behavior of the system is similar to the behavior of
electromagnetic pulsar. The derived spectral formula describes
the spectrum of photons generated by the Fermilab Tevatron. In case
that the particles in bunches are of the same charge as in LHC, then,
it is
necessary to replace the function sine by cosine in the final
spectral formula. Now let us approach the
theory and explicit calculation of the spectrum.

This process is the generalization of the one-charge synergic
synchrotron-\v{C}erenkov
radiation which has been calculated in source theory two decades ago
by Schwinger et al. (1976). We will follow the Schwinger article and
also the author articles
(Pardy, 1994d, 2000, 2002) as the starting point.
Although our final problem is the radiation of the two-charge system in vacuum,
we consider, first in general, the presence of dielectric medium, which is
represented by the
phenomenological index of refraction $n$ and it is well known that this
phenomenological constant depends on the external magnetic field.
Introducing the phenomenological constant enables to consider
also the \v{C}erenkovian processes. Later we put $n = 1$.

We will investigate here how the original Schwinger (et al.) spectral
formula of the synergic c synchrotron \v{C}erenkov radiation of the
charged particle is modified if we consider the electron and positron
moving at the opposite  angular velocities.
This problem is an analogue of the linear (Pardy, 1997) and circular problem solved
by author (Pardy, 2000). We will show that the
original spectral formula
of the synergic synchrotron-\v Cerenkov radiation
is modulated by function
$4\sin^{2}(\omega t)$ where $\omega$ is the frequency of
the synergistic radiation produced by the system and it does not depend on
the orbital angular frequency of electron or positron.
We will use here the fundamental ingredients of Schwinger
source theory (Schwinger, 1970, 1973; Dittrich, 1978; Pardy, 1994c, d, e)
to determine the power spectral formula.

\subsection{Formulation of the electromagnetic problem}

The basic formula of the Schwinger source theory is the so called
vacuum to vacuum amplitude:$\langle 0_{+}|0_{-} \rangle = \exp\{\frac{i}{\hbar}\*W\},$
where in case of the electromagnetic field in the medium, the action
$W$ is given by the following formula:
$$W = \frac{1}{2c^2}\*\int\,(dx)(dx')J^{\mu}(x){D}_{+\mu\nu}(x-x')J^{\nu}(x'),\eqno(63)$$
where

$${D}_{+}^{\mu\nu} = \frac{\mu}{c}[g^{\mu\nu} +
(1-n^{-2})\beta^{\mu}\beta^{\nu}]\*{D}_{+}(x-x'),\eqno(64)$$
where $\beta^{\mu}\, \equiv \, (1,{\bf 0})$, $J^{\mu}\, \equiv \,(c\varrho,{\bf
J})$ is the conserved current, $\mu$ is the magnetic permeability of
the medium, $\epsilon$ is the dielectric constant od the medium and
$n=\sqrt{\epsilon\mu}$ is the index of refraction of the medium.
Function ${D}_{+}$ is defined as in eq. (10) (Schwinger et al., 1976):

$$D_{+}(x-x') =\frac {i}{4\pi^2\*c}\*\int_{0}^{\infty}d\omega
\frac {\sin\frac{n\omega}{c}|{\bf x}-{\bf x}'|}{|{\bf x} - {\bf x}'|}\*
e^{-i\omega|t-t'|}.\eqno(65)$$

The probability of the persistence of vacuum follows from the vacuum
amplitude (1) where ${\rm Im}\;W$ is the basis for the following definition of the spectral
function $P(\omega,t)$:

$$-\frac{2}{\hbar}\*{\rm Im}\;W \;\stackrel{d}{=} \; -\,
\int\,dtd\omega\frac{P(\omega,t)}{\hbar\omega}.\eqno(66) $$

Now, if we insert eq. (64) into eq. (66), we get
after extracting $P(\omega,t)$ the following general expression
for this spectral function:

$$P(\omega,t) = -\frac{\omega}{4\pi^2}\*\frac{\mu}{n^2}\*\int\,d{\bf x}
d{\bf x}'dt'\left[\frac{\sin\frac{n\omega}{c}|{\bf x} -
{\bf x}'|}{|{\bf x} - {\bf x}'|}\right]$$

$$\times \quad \cos[\omega\*(t-t')]\*[\varrho({\bf x},t)\varrho({\bf x}',t')
- \frac{n^2}{c^2}\*{\bf J}({\bf x},t)\cdot{\bf J}({\bf x}',t')],\eqno(67)$$
which is an analogue of the formula (11).

Let us recall that the last formula can be derived also in the classical
electrodynamic context as it is shown for instance in the
Schwinger article (Schwinger, 1949).
The derivation of the power spectral formula from the
vacuum amplitude is more simple.

\subsection{The radiation of two opposite charges}

Now, we will apply the formula (67) to the two-body system
with the opposite charges moving at the opposite angular velocities
in order to get in general synergic synchrotron-\v Cerenkov
radiation of electron and positron moving in a uniform
magnetic field

While the synchrotron radiation is generated in a vacuum, the synergic
synchrotron-\v Cerenkov radiation can produced only in a
medium with dielectric constant $n$.
We suppose the circular motion with velocity ${\bf v}$
in the plane perpendicular to the
direction of the constant magnetic field ${\bf H}$ (chosen to be
in the $+z$ direction).

We can write the following formulas for the
charge density $\varrho$ and for the current
density ${\bf J}$ of the two-body system with opposite charges and opposite
angular velocities:

$$
\varrho({\bf x},t) = e\*\delta\*({\bf x}-{\bf x_{1}}(t))
-e\*\delta\*({\bf x}-{\bf x_{2}}(t))\eqno(68)$$
and

$$
{\bf J}({\bf x},t) = e\*{\bf v}_{1}(t)\*\delta\*({\bf x}-{\bf x_{1}}(t))
-e\*{\bf v}_{2}(t)\*\delta\*({\bf x}-{\bf x_{2}}(t))\eqno(69)$$
with

$$
{\bf x}_{1}(t)  = {\bf x}(t) =
R({\bf i}\cos(\omega_{0}t) + {\bf j}\sin(\omega_{0}t)),\eqno(70)$$

$$
{\bf x}_{2}(t) =
R({\bf i}\cos(-\omega_{0}t) +
{\bf j}\sin(-\omega_{0}t) =
{\bf x}(-\omega_{0},t) = {\bf x}(-t).\eqno(71)$$

The absolute values of velocities of both particles are
the same, or $|{\bf v}_{1}(t)| =  |{\bf v}_{2}(t)| = v$, where
($H = |{\bf H}|, E =$ energy of a particle)

$$
{\bf v}(t) = d{\bf x}/dt, \hspace{5mm} \omega_{0} = v/R, \hspace{5mm}
R = \frac {\beta\*E}{eH}, \hspace{5mm}
\beta = v/c, \hspace{5mm} v = |{\bf v}|.\eqno(72)$$

After insertion of eqs. (68)--(71) into eq. (67), and after some mathematical
operations we get

$$P(\omega,t) =
-\frac{\omega}{4\pi^2}\*\frac{\mu}{n^2}e^{2}\*\int_{-\infty}^{\infty}\,
dt'\cos(t-t')\sum_{i,j = 1}^{2}(-1)^{i+j}
$$

$$\times \quad \left[1 - \frac {{\bf v}_{i}(t)\cdot {\bf v}_{j}(t')}{c^{2}}n^{2}\right]
\left\{\frac{\sin\frac {n\omega}{c}|{\bf x}_{i}(t) -{\bf x}_{j}(t')|}
{|{\bf x}_{i}(t) -{\bf x}_{j}(t')|}\right\}.\eqno(73)$$

Let us remark, that for situation of the identical charges, the factor
$(-1)^{i + j}$ must be replaced by 1.

Using $t' = t + \tau$, we get for

$$
{\bf x}_{i}(t) -{\bf x}_{j}(t')
\stackrel{d}{=} {\bf A}_{ij},\eqno(74)$$

$$
|{\bf A}_{ij}| = [R^{2} + R^{2} - 2RR\cos(\omega_{0}\tau +
\alpha_{ij})]^{1/2} =
2R\left|\sin\left(\frac {\omega_{0}\tau +
      \alpha_{ij}}{2}\right)\right|,
\eqno(75)$$
where $\alpha_{ij}$ were evaluated as follows:

$$
\alpha_{11} = 0,\quad \alpha_{12 } = 2\omega_{0}t,
\quad \alpha_{21} = 2\omega_{0}t, \quad  \alpha_{22} = 0.\eqno(76)$$

Using

$$
{\bf v}_{i}(t)\cdot{}{\bf v}_{j}(t+\tau) = \omega_{0}^{2}R^{2}
\cos(\omega_{0}\tau +
\alpha_{ij}),\eqno(77)$$
and relation (75) we get with $v= \omega_{0}R$

$$P(\omega,t) =
-\frac{\omega}{4\pi^2}\*\frac{\mu}{n^2}e^{2}\*\int_{-\infty}^{\infty}\,
d\tau \cos\omega\tau \sum_{i,j = 1}^{2}(-1)^{i+j} $$

$$\times \quad
\left[1 - \frac {n^{2}}{c^{2}}v^{2}\cos(\omega_{0}\tau + \alpha_{ij})\right]
\left\{\frac{\sin\left[\frac {2Rn\omega}{c}
\sin\left(\frac {(\omega_{0}\tau + \alpha_{ij})}
{2}\right)\right]}
{2R\sin\left(\frac {(\omega_{0}\tau +
      \alpha_{ij})}{2}\right)}\right\}.
\eqno(78)$$

Introducing new variable $T$ by relation

$$
\omega_{0}\tau + \alpha_{ij} = \omega_{0}T\eqno(79)$$
for every integral in eq. (78),
we get $P(\omega,t)$ in the following form

$$P(\omega,t) =
-\frac{\omega}{4\pi^2}\frac {e^{2}}{2R}
\*\frac{\mu}{n^2}\*\int_{-\infty}^{\infty} dT \sum_{i,j=1}^{2}(-1)^{i+j}
 $$

$$\times\quad
\cos(\omega T - \frac {\omega}{\omega_{0}}\alpha_{ij})
\left[1 - \frac {c^{2}}{n^{2}}v^{2}\cos(\omega_{0} T \right]
\left\{\frac{\sin\left[\frac {2Rn\omega}{c}\sin
\left(\frac {\omega_{0}T}{2}\right)\right]}
{\sin\left(\frac {\omega_{0}T}{2}\right)}\right\}.\eqno(80)$$

The last formula can be written in the more compact form,

$$P(\omega,t) = -\frac {\omega}{4\pi^{2}}\frac {\mu}{n^{2}}\frac {e^{2}}{2R}
\sum_{i,j=1}^{2}(-1)^{i+j}\left\{P_{1}^{(ij)} -\frac {n^{2}}{c^{2}}v^{2}
P_{2}^{(ij)}\right\},\eqno(81)$$
where

$${P}^{(ij)} = J_{1a}^{(ij)}\cos\frac {\omega}{\omega_{0}}\alpha_{ij} +
J_{1b}^{(ij)}\sin\frac {\omega}{\omega_{0}}\alpha_{ij}\eqno(82)$$
and

$$
P_{2}^{(ij)} = J_{2A}^{(ij)}
\cos\frac {\omega}{\omega_{0}}\alpha_{ij} +
J_{2B}^{(ij)}\sin\frac {\omega}{\omega_{0}}\alpha_{ij},\eqno(83)$$
where

$$
J_{1a}^{(ij)} = \int_{-\infty}^{\infty}dT\cos\omega T
\left\{\frac{\sin\left[\frac {2Rn\omega}{c}\sin
\left(\frac {\omega_{0}T}{2}\right)\right]}
{\sin\left(\frac {\omega_{0}T}{2}\right)}\right\},\eqno(84)$$

$$
J_{1b}^{(ij)} = \int_{-\infty}^{\infty}dT\sin\omega T
\left\{\frac{\sin\left[\frac {2Rn\omega}{c}\sin
\left(\frac {\omega_{0}T}{2}\right)\right]}
{\sin\left(\frac {\omega_{0}T}{2}\right)}\right\},\eqno(85)$$

$$
J_{2A}^{(ij)} = \int_{-\infty}^{\infty}dT\cos\omega_{0}T\cos\omega T
\left\{\frac{\sin\left[\frac {2Rn\omega}{c}\sin
\left(\frac {\omega_{0}T}{2}\right)\right]}
{\sin\left(\frac {\omega_{0}T}{2}\right)}\right\},\eqno(86)
$$

$$
J_{2B}^{(ij)} = \int_{-\infty}^{\infty}dT\cos\omega_{0}T\sin\omega T
\left\{\frac{\sin\left[\frac {2Rn\omega}{c}\sin
\left(\frac {\omega_{0}T}{2}\right)\right]}
{\sin\left(\frac {\omega_{0}T}{2}\right)}\right\},\eqno(87)$$

Using

$$
\omega_{0}T = \varphi + 2\pi\*l,  \hspace{7mm} \varphi\in(-\pi,\pi),\;
\quad l = 0,\, \pm1,\, \pm2,\, ... ,\eqno(88)
$$
we can transform the $T$-integral into the sum of the telescopic
integrals according to the scheme:

$$
\int_{-\infty}^{\infty}dT\quad\longrightarrow \quad\frac {1}{\omega_{0}}
\sum_{l = -\infty}^{l = \infty}\int_{-\pi}^{\pi}d\varphi.\eqno(89)$$

Using the fact that for the odd functions $f(\varphi)$ and $g(l)$,
the relations are valid

$$
\int_{-\pi}^{\pi}f(\varphi)d\varphi = 0; \quad \sum_{l=-\infty}^{l =
\infty}g(l) = 0,\eqno(90)$$
we can write

$$
J_{1a}^{(ij)} = \frac {1}{\omega_{0}}\sum_{l}\int_{-\pi}^{\pi}
d\varphi\left\{\cos{\frac {\omega}{\omega_{0}}\varphi\cos{2\pi l}
\frac{\omega}{\omega_{0}}}\right\}
\left\{\frac{\sin\left[\frac {2Rn\omega}{c}\sin
\left(\frac {\varphi}{2}\right)\right]}
{\sin\left(\frac {\varphi}{2}\right)}\right\},\eqno(91)$$

$$
J_{1b}^{(ij)} = 0.\eqno(92)$$

For integrals with indices A, B we get:

$$
J_{2A}^{(ij)} = \frac {1}{\omega_{0}}\sum_{l}\int_{-\pi}^{\pi}
d\varphi\cos\varphi
\left\{\cos{\frac {\omega}{\omega_{0}}\varphi\cos{2\pi l}
\frac{\omega}{\omega_{0}}}\right\}
\left\{\frac{\sin\left[\frac {2Rn\omega}{c}\sin
\left(\frac {\varphi}{2}\right)\right]}
{\sin\left(\frac {\varphi}{2}\right)}\right\},\eqno(93)$$

$$
J_{2B}^{(ij)} = 0,\eqno(94)$$

So, the power spectral formula (80) is of the form:

$$
P(\omega,t) = -\frac {\omega}{4\pi^{2}}\frac {\mu}{n^{2}}\frac {e^{2}}{2R}
\sum_{i,j=1}^{2}(-1)^{i+j}\left\{P_{1}^{(ij)} - n^{2}\beta^{2}
P_{2}^{(ij)}\right\};\quad \beta = \frac {v}{c},\eqno(95)$$
where

$$
P_{1}^{(ij)} = J_{1a}^{(ij)}\cos\frac {\omega}{\omega_{0}}\alpha_{ij}
\eqno(96)$$
and

$$P_{2}^{(ij)} = J_{2A}^{(ij)}\cos\frac
{\omega}{\omega_{0}}\alpha_{ij}.
\eqno(97)$$
Using the Poisson theorem

$$
\sum_{l = -\infty}^{\infty}\cos 2\pi\frac {\omega}{\omega_{0}}l  =
\sum_{k=-\infty}^{\infty}\omega_{0}\delta(\omega - \omega_{0}l),\eqno(98)$$
the definition of the Bessel functions $J_{2l}$ and their
corresponding derivations and integrals

$$
\frac {1}{2\pi}\int_{-\pi}^{\pi}d\varphi\cos\left(z\sin\frac {\varphi}{2}
\right)\cos l\varphi  = J_{2l}(z),\eqno(99)$$

$$
\frac {1}{2\pi}\int_{-\pi}^{\pi}d\varphi\sin\left(z\sin\frac {\varphi}{2}
\right)\sin(\varphi/2)\cos l\varphi  = - J'_{2l}(z),\eqno(100)$$

$$
\frac {1}{2\pi}\int_{-\pi}^{\pi}d\varphi
\frac{\sin\left(z\sin\frac{\varphi}{2}\right)}
{\sin(\varphi/2)}\cos l\varphi  = \int_{0}^{z}J_{2l}(x)dx,\eqno(101)$$
and using  equations

$$
\sum_{i,j = 1}^{2}(-1)^{i+j}\cos\frac {\omega}{\omega_{0}}\alpha_{ij}
= 2(1-\cos 2\omega t) = 4\sin^{2}\omega t,\eqno(102)$$

we get with the definition of the partial power spectrum $P_{l}$

$$
P(\omega) = \sum_{l=1}^{\infty}  \delta(\omega - l\omega_{0})P_{l},\eqno(103)$$
the following final form of the partial power spectrum
generated by motion of
two-charge system moving in the cyclotron:

$$
P_{l}(\omega,t) = [4(\sin\omega t)^{2}]
\frac {e^2}{\pi\*n^2}\*\frac {\omega\mu\omega_{0}}{v}\*
\left(2n^2\beta^2J'_{2l}(2ln\beta) -
(1 - n^2\*\beta^2)\*\int_{0}^{2ln\beta}dxJ_{2l}(x)\right).\eqno(104)$$

So we see that the spectrum generated by the system of electron and positron
is formed in such a way that the original synchrotron spectrum generated by
electron is modulated by function
$4\sin^{2}(\omega t)$. The derived formula
involves also the synergic process composed from
the synchrotron radiation and the
\v Cerenkov radiation for electron velocity $v > c/n$ in a medium.

Our  goal is to apply the last formula
in situation where there is a vacuum. In
this case we can put $\mu = 1, n = 1$ in the last formula and so we have

$$P_{l}(\omega,t) =
4 \sin^{2}\left(\omega t\right)
\frac {e^2}{\pi}\*\frac {\omega\omega_{0}}{v}\*
\left(2\beta^2J'_{2l}(2l\beta) -
(1 - \beta^2)\*\int_{0}^{2l\beta}dxJ_{2l}(x)\right).\eqno(105)$$

So, we see, that final formula describing the
opposite motion of electron and positron in accelerator is
of the form

$$
P_{l,pair}(\omega,t) =
4 \sin^{2}\left(\omega t\right)P_{l(electron)}\left(\omega\right),\eqno(106)$$
where $P_{electron}$ is the spectrum of radiation only of
electron. For the same charges it is necessary to replace sine by
cosine in the final formula.

The result (106)is surprising because we naively
expected that the total radiation of the opposite charges should be

$$P_{l}(\omega,t) =
P_{l(electron)}\left(\omega, t \right) +
P_{l(positron)}\left(\omega, t \right).\eqno(107)$$

So, we see that the resulting radiation  can not be considered
as generated by the isolated particles but by a synergic production of
a system of particles and magnetic field. At the same time we cannot
interpret the result as a result of interference of two sources
because the distance between sources radically changes and so, the
condition of an interference is not fulfilled.

The classical electrodynamics formula (106) changes our naive opinion
on the electrodynamic  processes in the magnetic
field. From the last formula  it follows that at time $t = \pi k/\omega $
there is no radiation of the frequency $\omega$.
The spectrum oscillates with frequency $\omega$.
If the radiation were generated not in the synergic way,  then
the spectral formula would be composed from two parts corresponding
to two isolated sources.

\section{The two center circular motions}

The situation which we have analyzed was the ideal situation where the angle of collision of positron and electron
was equal to $\pi$. Now, the question arises what is the modification of a spectral formula when the collision angle between
particles differs from $\pi$.
It can be easily seen that if the second particle follows the shifted circle trajectory, then the collision angle
differs from $\pi$.  Let us suppose that the center of the circular
trajectory of the second particle
has coordinates $(a,0)$. It can be easy to
see from the geometry of the situation and from the plane geometry
that the collision angle is $ \pi - \alpha, \alpha
 \approx \tan \alpha \approx a/R$ where
$R$ is a radius of the first or second circle. The same result follows
from the analytical geometry of the situation.

While the equation of the first particle is the equation of the original trajectory, or this is eq. (70)

$$
{\bf x}_{1}(t) = {\bf x}(t) = R({\bf i}\cos(\omega_0 t) + {\bf j}\sin(\omega_0 t)),\eqno(108)$$
the equation of a circle with a shifted center is as follows:

$$
{\bf x}_{2}(t) = {\bf x}(t) = R({\bf i}(\frac {a}{R}+\cos(-\omega_0 t)) +
{\bf j}\sin(-\omega_0 t)) = {\bf x(-t)} + {\bf i}a.\eqno(109)$$

The absolute values of velocities of both particles are equal and the
relation (72) is valid.
Instead of equation (74) we have for radius vectors of particle trajectories:

$$
{\bf x}_{i}(t) -{\bf x}_{j}(t')
\stackrel{d}{=} {\bf B}_{ij},\eqno(110)$$
where ${\bf B}_{11} = {\bf A}_{11}, {\bf B}_{12} = {\bf A}_{12} -
{\bf i}a, {\bf B}_{21} = {\bf A}_{21} + {\bf i}a,
{\bf B}_{22} = {\bf A}_{22}$.

In general, we can write the last information on
coefficients ${\bf B}_{ij}$ as follows:

$$
 {\bf B}_{ij} = {\bf A}_{ij} + \varepsilon_{ij}{\bf i}a ,\eqno(111)$$
where $\varepsilon_{11} = 0, \varepsilon_{12} = -1, \varepsilon_{21} = 1,
\varepsilon_{22} = 0.$

For motion of particles along trajectories the absolute value of vector
${\bf A}_{ij} \gg a$ during the most part of the trajectory.
It means, we can determine
$B_{ij}$ approximatively. After elementary operations, we get:

$$
|{\bf B_{ij}}| =  (A_{ij}^{2} +
2|{\bf A}_{ij}|\varepsilon_{ij}a\cos\varphi_{ij} +
a^{2}\varepsilon_{ij}^{2})^{1/2},\eqno(112)$$
where $\cos\varphi_{ij}$ can be expressed by the $x$-component of
vector ${\bf A}_{ij}$ and $|{\bf A}_{ij}|$ as follows:

$$
\cos\varphi_{ij} =  \frac{(A_{ij})_{x}}{|{\bf A}_{ij}|}.  \eqno(113)$$

After elementary trigonometric operations, we derive the following
formula  for $(A_{ij})_{x}$:

$$
(A_{ij})_{x} = 2R\sin\frac {2\omega_{0}t + \omega_{0}\tau}{2}
\sin\frac {\omega_{0}\tau}{2}.  \eqno(114)$$

Then, using equation (114), we get with $\varepsilon = a/R$

$$|{\bf B_{ij}}| = 2R (\sin^{2}\frac {\omega_{0}\tau + \alpha_{ij}}{2}
+ \varepsilon\varepsilon_{ij}\sin\frac {2\omega_{0}t + \omega_{0}\tau}{2}
\sin\frac {\omega_{0}\tau}{2} +
\varepsilon^{2}\frac{\varepsilon_{ij}^{2}}{4})^{1/2}.\eqno(115)$$

In order to perform the $\tau$-integration the substitution must be
introduced. However, the substitution $\omega_{0}\tau + \alpha_{ij} =
\omega_{0}T$ does not work. So we define the substitution $\tau = \tau(T)$
by the following transcendental equation (we neglect the term with
$\varepsilon^{2}$):

$$\left[\sin^{2}\frac {\omega_{0}\tau + \alpha_{ij}}{2}
+ \varepsilon\varepsilon_{ij}\sin\frac {\omega_{0}t + \omega_{0}\tau}{2}
\sin\frac {\omega_{0}\tau}{2}\right]^{1/2} =
\sin\frac {\omega_{0}T}{2}.  \eqno(116)$$

Or, after some trigonometrical modifications  and using
the approximative formula $(1 + x)^{1/2} \approx 1 + x/2$ for $x\ll 1$

$$
\left[ ./. \right]^{1/2} \approx
\sin \left(\frac {\omega_{0}\tau + \alpha_{ij}}{2}\right)
+ \frac{\varepsilon}{2}\varepsilon_{ij}
\sin\left(\frac {2\omega_{0}\tau + 2\omega_{0}t -\alpha_{ij}}{2}\right)
=  \sin\frac {\omega_{0}T}{2}.  \eqno(117)$$

We se that for $\varepsilon = 0$ the substitution is
$\omega_{0}\tau + \alpha_{ij} = \omega_{0}T$. The equation (117) is the
transcendental  equation and the exact solution is the function
$\tau = \tau(T)$. We are looking for the solution of equation (117) in the
approximative form using the approximation $\sin x \approx x$.

Then, instead of (117) we have:

$$ \left(\frac {\omega_{0}\tau + \alpha_{ij}}{2}\right)
+ \frac{\varepsilon}{2}\varepsilon_{ij}
\left(\frac {2\omega_{0}\tau + 2\omega_{0}t -\alpha_{ij}}{2}\right)
= \frac {\omega_{0}T}{2}\eqno(118)$$
Using substitution

$$\omega_{0}\tau + \alpha_{ij} = \omega_{0}T + \omega_{0}\varepsilon A
\eqno(119)$$
in eq. (118) we get, to the first order in $\varepsilon$-term:

$$
A = -\frac{\varepsilon_{ij}}{2\omega_{0}}(\omega_{0}T - 2\alpha_{ij} +
2\omega_{0}t).  \eqno(120)$$

Then, after some algebraic manipulation we get:

$$
\omega_{0}\tau  +  \alpha_{ij} =
\omega_{0}T(1 -\frac{\varepsilon}{2} \varepsilon_{ij})
- \varepsilon\varepsilon_{ij}\omega_{0}t(-1)^{i+j}\eqno(121)$$
and

$$
\omega\tau  = \omega T (1 - \frac{\varepsilon}{2}\varepsilon_{ij}) -
\frac{\omega}{\omega_{0}}
\left(\varepsilon\varepsilon_{ij}(-1)^{i+j}\omega_{0}t +
\alpha_{ij}\right).\eqno(122)$$

For small time $t$, we can write approximately:

$$\cos(\omega_{0}\tau + \alpha_{ij})
\approx \cos\omega_{0} T(1 -
\frac{\varepsilon}{2}\varepsilon_{ij})\eqno(123)$$
and from eq. (122)

$$
d\tau = dT(1 - \frac{\varepsilon}{2}\varepsilon_{ij}).  \eqno(124)$$

So, in case of the eccentric circles
the formula (118) can be obtained from non-perturbative formula (80)
only by transformation

$$
T \quad \longrightarrow \quad T(1 - \frac{\varepsilon}{2}\varepsilon_{ij}) ;
\quad \alpha_{ij} \quad\longrightarrow \quad
\left(\varepsilon\varepsilon_{ij}(-1)^{i+j}\omega_{0}t + \alpha_{ij}\right)
= \tilde {\alpha}_{ij}, \eqno(125)$$
excepting specific term involving sine functions.

Then, instead of formula (80) we get:

$$P(\omega,t) =
-\frac{\omega}{4\pi^2}\frac {e^{2}}{2R}
\*\frac{\mu}{n^2}\*\int_{-\infty}^{\infty} dT \sum_{i,j=1}^{2}(-1)^{i+j}$$

$$\times \quad
\cos(\omega T - \frac {\omega}{\omega_{0}}\tilde{\alpha}_{ij})
\left[1 - \frac {c^{2}}{n^{2}}v^{2}
\cos(\omega_{0} T)
\right]
\left\{\frac{\sin\left[\frac {2Rn\omega}{c}\sin
\left(\frac {\omega_{0}{\tilde T}}{2}\right)\right]}
{\sin\left(\frac {\omega_{0}{\tilde T}}{2}\right)}\right\}.\eqno(126)$$
where $\tilde T = T(1 - \frac {\varepsilon}{2}\varepsilon_{ij})$.
We see that only $\tilde T$  and the $\alpha$ term
are the new modification of the original formula (80).

However, because $\varepsilon$ term in the sine functions
is of very small influence on the behavior of the total  function for
finite time $t$, we can neglect it and write approximatively:

$$P(\omega,t) =
-\frac{\omega}{4\pi^2}\frac {e^{2}}{2R}
\*\frac{\mu}{n^2}\*\int_{-\infty}^{\infty} dT \sum_{i,j=1}^{2}(-1)^{i+j}
 $$

$$\times\quad
\cos(\omega T - \frac {\omega}{\omega_{0}}\tilde{\alpha}_{ij})
\left[1 - \frac {c^{2}}{n^{2}}v^{2}\cos(\omega_{0} T)
\right]
\left\{\frac{\sin\left[\frac {2Rn\omega}{c}\sin
\left(\frac {\omega_{0}T}{2}\right)\right]}
{\sin\left(\frac {\omega_{0}T}{2}\right)}\right\}.\eqno(127)$$

So, we se that only difference with the original radiation formula is in
variable  ${\tilde \alpha}_{ij}$.

It means that instead of sum (102) we have the following sum:

$$
\sum_{i,j = 1}^{2}(-1)^{i+j}\cos\frac {\omega}{\omega_{0}}
{\tilde\alpha_{ij}} =
2(1-\cos 2\omega t \cos\varepsilon\omega t).  \eqno(128)$$

It means that the one electron radiation formula is not
modulated by $[\sin\omega t]^{2}$
but by the formula (128) and the final formula of for the power spectrum
is as follows:

$$
P_{l}(\omega,t) =
2(1-\cos 2\omega t \cos\varepsilon\omega t)
P_{l(electron)}\left(\omega\right).  \eqno(129)$$

For $\varepsilon \to 0$, we get the original formula (106).

\section{SUMMARY AND DISCUSSION}

We have derived, in the first part of the article,
the total quantum loss of energy of the binary.
The energy loss is caused by the emission of gravitons during the
motion of the two binary bodies
around each other under their gravitational interaction.
The energy-loss formulas of the production of gravitons
are derived here in the source theory.
It is evident that the
production of gravitons by the binary system is not homogenous and
isotropical in space.
So, the binary forms the ``gravitational light house'' where instead
of the light
photons of the electromagnetic pulsar are
 the gravitons. The detector of the
gravitational waves evidently detects the gravitational pulses.

This section  is an extended and revised
version of the older author's article (Pardy, 1983a) and preprints
(Pardy,1994a,b), in which only the spectral
formulas were derived. Here, in the first part of the article,
we have derived the quantum energy-loss
formulas for the linear gravitational field. Linear field corresponds to the
weak field limit of the Einstein gravity.

The power spectrum formulas  involving radiative corrections
are derived in  the following part of this article, also in the
framework of the source theory.
The general relativity necessarily does not contain the method how
to express the quantum effects together with the radiative corrections
by the geometrical language.
So, it cannot give the answer on the
production of gravitons and on the graviton propagator with radiative
corrections.
This section therefore deals with the quantum energy loss caused by the
production of gravitons and by the radiative corrections in the graviton
propagator in case of the motion of a binary.

We believe the situation in the gravity problems with radiative corrections
is similar to the QED situation many years
ago when the QED radiative corrections were theoretically predicted
and then experimentally
confirmed for instance in case of he Lamb shift, or, of the anomalous
magnetic moment of electron.

Astrophysics is, in a crucial position in proving the influence
of radiative corrections on the dynamics in the cosmic space.
We hope that the further astrophysical observations
will confirm the quantum version of the energy loss of the binary
with graviton propagator with radiative corrections.

In the last part of this article on pulsars we have derived  the
power spectrum formula
of the synchrotron radiation generated by the
electron and positron moving at the opposite angular
velocities in homogeneous magnetic field. It forms an analogue of the author
article (Pardy, 1997)
where only comoving electrons, or positrons was considered,
and it forms the modified
author preprints  (Pardy, 2000a; 2001) and articles (Pardy, 200b; 2002,)
where the power spectrum is calculated for two charges
performing the retrograde motion in a magnetic field. The frequency of
motion was the same because the diameter of
the circle was considered the same for both charges.
The retrograde motion with different diameters was not considered.

It is surprising that the spectrum
depends periodically on radiation frequency $\omega$ and time which means
that the system composed from electron, positron and magnetic
field behaves as a pulsating system. While such pulsar can be represented by a
terrestrial experimental arrangement it is possible to consider also the
cosmological existence in some modified conditions.

To our knowledge, our result is not involved in the
classical monographs on the electromagnetic theory and at the same time
it was not still studied by the
accelerator experts investigating the synchrotron radiation of bunches.
This effect was not described in textbooks on
classical electromagnetic field and on the synchrotron radiation.
We hope that sooner or later this effect will be verified
by the accelerator physicists.
The radiative corrections
obviously influence the synergistic spectrum of photons
(Pardy, 1994c,d).

The particle laboratories used instead of the single electron and positron the
bunches with 10$^{10}$ electrons or positrons  in one bunch of
volume 300$ \mu$m $\times$ 40$ \mu$m $\times$ 0.01 m.
So, in some approximation we can replace the charge of electron
and positron by the charges {Q} and {-Q} of both bunches in order to
get the realistic intensity of photons. Nevertheless the synergic character of
the radiation of two bunches moving at the opposite
direction in a magnetic field is conserved.

\vspace{7mm}
\noindent
{\bf REFERENCES}

\vspace{7mm}
\noindent
Damour, T.; Taylor, J. H.  Phys. Rev. {\bf D 45} No. 6 1868 (1992). \\
Dittrich, W.  Fortschritte der Physik {\bf 26}  289 (1978).\\
Evans, L. R. {\it The Large Hadron Collider - Present Status and Prospects}
CERN-OPEN-99-332  CERN Geneva (1999).\\
Gold, T. Nature {\bf 218} 731 (1968).\\
Goldreich, P.;  Julian, W. H. Astrophys. J. {\bf 157} 869 (1969).\\
Graham-Smith, F.  Rep. Prog. Phys. {\bf 66} 173 (2003). \\
Hewish, A.; Bell, S. J.; Pilkington, J. D. H. ; et al. Nature {\bf 217} 709
(1968).\\
Huguenin, G. R; Taylor, J. H.;  Goad, L. E.; Hartai, A.;  Orsten, G. S. F.; Rodman, A. K.
Nature {\bf 219} 576 (1968).\\
Hulse, R. A.; Taylor, J. H. Astrophys. J. Lett., {\bf 195} L51-L53 (1975).\\
Cho C. F. and Harri Dass, N. D. Ann. Phys. (NY) {\bf 90} 406 (1976).\\
Landau, L. D. Phys. Zs. Sowjet {\bf 1} 285 (1932).\\
Manchester, R. N.  Phil. Trans. R. Soc. Lond. {\bf A 341} 3 (1992). \\
Manoukian, E. B.  GRG {\bf 22} 501 (1990). \\
Melrose, D. B. Phil. Trans. R. Soc. Lond. {\bf A 341} 105 (1992).\\
Pardy, M. GRG {\bf 15} No. 11 1027 (1983a).\\
Pardy, M. Phys. Lett. {\bf 94A} 30 No. 1 (1983b).\\
Pardy, M. Phys. Lett. {\bf 140A} 51 Nos. 1,2 (1989).\\
Pardy, M. CERN-TH.7239/94 (1994a). \\
Pardy, M. CERN-TH.7299/94 (1994b). \\
Pardy, M. Phys. Lett. {\bf B 325} 517 (1994c).\\
Pardy, M. Phys. Lett. {\bf A 189} 227 (1994d). \\
Pardy, M. Phys. Lett. {\bf B 336} 362 (1994e). \\
Pardy, M. Phys. Rev. {\bf A 55} No. 3 1647 (1997).\\
Pardy, M. hep-ph/0001277 (2000a).\\
Pardy, M. Int. Journal of Theor. Phys. {\bf 39} No. 4 1109 (2000b).\\
Pardy, M. hep-ph/011036 (2001). \\
Pardy, M.  Int. Journal of Theor. Phys. {\bf 41} No. 6 1155 (2002).\\
Seiradakis, J. H.; Wielebinski, R. Astronomy \& Astrophysics Review
manuscript (2004), hep-ph/0410022 (2004).\\
Schott, G. A. {\it Electromagnetic Radiation} (Cambridge University
Press, 1912).\\
Schwinger, J. Phys. Rev. {\bf 75} 1912 (1949). \\
Schwinger, J.  GRG {\bf 7} No. 3  251 (1976).\\
Schwinger, J. {\it Particles, Sources and Fields}, Vol. I,
(Addison-Wesley, Reading, Mass., 1970). \\
Schwinger, J. {\it Particles, Sources and Fields}, Vol. II,
(Addison-Wesley, Reading, Mass., 1973). \\
Schwinger, J. Phys. Rev. {\bf 75} 1912 (1949). \\
Schwinger, J.; Tsai, W. Y; Erber, T. Ann. Phys. (NY) {\bf 96} 303 (1976).\\
Sokolov, A. A.;  Ternov, I. M. The relativistic electron (Moscow, Nauka, 1983).
(in Russian). \\
Taylor, J. H.; Wolszczan, A.; Damour, T.;  Weisberg, J. M. Nature
{\bf 355} 132 (1992). \\
Taylor, J. H. Jr. Binary Pulsars and Relativistic Gravity
(Nobel Lecture, 1993).\\
Weinberg,  S.  {\it Gravitation and Cosmology} (John Wiley and Sons, Inc.,
New York, 1972).
\end{document}